\documentclass[floatfix,twocolumn,amsmath,amssymb,aps,prb,showpacs,letter,superscriptaddress]{revtex4}

\usepackage{graphicx}
\usepackage{epsfig}
\usepackage{rotating}
\usepackage{amsmath}
\usepackage{amsfonts}
\usepackage{amssymb}
\usepackage{enumerate}
\usepackage{longtable}
\usepackage{dcolumn}
\usepackage{ulem}
\usepackage{bm}
\usepackage{color}

\definecolor{pink}{rgb}{0.855,0,0.627}

\begin{document}

\title{Thermodynamic properties of the Yb$_2$Ti$_2$O$_7$ pyrochlore as a function of temperature and magnetic field:
validation of a quantum spin ice exchange Hamiltonian}

\author{N. R. Hayre}
\affiliation{Physics Department, University of California at Davis, Davis, CA 95616}
\author{K. A. Ross}
\affiliation{Department of Physics and Astronomy, McMaster University, Hamilton, Ontario, L8S 4M1, Canada}
\affiliation{Institute for Quantum Matter and Department of Physics and Astronomy, Johns Hopkins University, Baltimore, Maryland 21218}
\affiliation{NIST Center for Neutron Research, National Institute of Standards and Technology, Gaithersburg, Maryland 20899}
\author{R. Applegate}
\affiliation{Physics Department, University of California at Davis, Davis, CA 95616}
\author{T. Lin}
\affiliation{Department of Physics and Astronomy, University of Waterloo, Waterloo, Ontario, N2L 3G1, Canada}
\author{R. R. P. Singh}
\affiliation{Physics Department, University of California at Davis, Davis, CA 95616}
\author{B. D. Gaulin}
\affiliation{Department of Physics and Astronomy, McMaster University, Hamilton, Ontario, L8S 4M1, Canada}
\affiliation{Brockhouse Institute for Materials Research, McMaster University, Hamilton, Ontario, L8S 4M1, Canada}
\affiliation{Canadian Institute for Advanced Research, 180 Dundas St. W., Toronto, Ontario, M5G 1Z8, Canada}
\author{M. J. P. Gingras}
\affiliation{Department of Physics and Astronomy, University of Waterloo, Waterloo, Ontario, N2L 3G1, Canada}
\affiliation{Canadian Institute for Advanced Research, 180 Dundas St. W., Toronto, Ontario, M5G 1Z8, Canada}
\affiliation{Perimeter Institute for Theoretical Physics, 31 Caroline North, Waterloo, Ontario, N2L-2Y5, Canada}

\date{\rm\today}

\begin{abstract}
The thermodynamic properties of the pyrochlore Yb$_2$Ti$_2$O$_7$ material are calculated using the numerical
linked-cluster (NLC) calculation method for an effective 
anisotropic-exchange spin-1/2 Hamiltonian
with parameters recently determined by fitting the neutron scattering spin wave data obtained at high magnetic field $h$. 
Magnetization, ${\bm M}(T,h)$, as a function of temperature $T$ and for different magnetic fields $h$
applied along the three high symmetry directions [100], [110] and [111], are 
compared with experimental measurements on the material for temperature $T > 1.8$ K.
The excellent agreement between experimentally measured and
calculated $M(T,h)$ over the entire temperature and magnetic field range considered
provides strong quantitative validation of the effective Hamiltonian.
It also confirms that fitting the high-field neutron spin wave spectra in the polarized paramagnetic state
is an excellent method for determining the microscopic exchange constants  of rare-earth insulating
magnets that are described by an effective spin-1/2  Hamiltonian.
Finally, we present results which demonstrate that a recent analysis
 of the polarized neutron scattering intensity of Yb$_2$Ti$_2$O$_7$ 
   using a random phase approximation (RPA) method  [Chang {\it et al.}, Nat. Comm. {\bf 3}, 992 (2012)]
does not provide a good description  of $M(T,h)$ for $T\lesssim 10$ K, that is in the entire
temperature regime where correlations become non-negligible.
\end{abstract}


\maketitle

\section{Introduction}

\subsection*{Magnetic rare-earth pyrochlore oxides and effective spin-1/2 quantum dynamics}

Quantum spin liquids are magnetic systems in which large quantum mechanical zero-point spin fluctuations
prevent the development of long-range order down to absolute zero temperature.
The search for real materials that display this phenomenology in two and three dimensions is
a very active research topic in the field of condensed matter physics.~\cite{balents}
The interest in spin liquids stems from the expectation that they may host non-trivial 
 quantum entanglement, topological order, as well as emergent fractionalized and deconfined low-energy excitations.
Spin liquids have also been conjectured as progenitors of unconventional superconductivity at
``high temperature''.
Highly frustrated magnets are  particularly apt at exhibiting large
quantum spin fluctuations. These magnets are realized in systems which consist of
 localized magnetic moments (spins) that reside on two and three dimensional lattices
of corner-sharing triangles or tetrahedra and which interact with effective antiferromagnetic
nearest-neighbor coupling.~\cite{effective_AF} 
Such highly frustrated magnets display an exponentially
large number of classical ground states.~\cite{moessner} 
This allows for quantum mechanical effects to 
be tremendously magnified compared to magnetic systems with conventional long-range magnetic order.~\cite{balents}

The pyrochlore oxides, of chemical formula RE$_2$M$_2$O$_7$,~\cite{gardner_RMP}
count a multitude of magnetic members.
In RE$_2$M$_2$O$_7$, RE is a magnetic trivalent (Lanthanide, 4f) 
rare-earth ion
(Gd$^{3+}$, Tb$^{3+}$, Dy$^{3+}$, Ho$^{3+}$, Er$^{3+}$, Yb$^{3+}$) or a non-magnetic 
ion (Y$^{3+}$, Lu$^{3+}$), 
while
M is  tetravalent,
typically a  transition metal ion,
 which can be either magnetic (Mo$^{4+}$, Mn$^{4+}$) or not 
(Ti$^{4+}$, Sn$^{4+}$, Zr$^{4+}$, Ge$^{4+}$).
Of relevance to the above discussion is the fact that the RE and M ions reside on two distinct and interpenetrating
three-dimensional lattices of corner-sharing tetrahedra. As a result, high geometric magnetic frustration ensues
whenever the RE-RE or M-M interaction is effectively antiferromagnetic.~\cite{effective_AF}
Among the RE$_2$M$_2$O$_7$ family, the magnetic rare-earth oxides, in which 
RE is magnetic and M is not, have been rather extensively studied.~\cite{gardner_RMP} 
These have revealed a number of fascinating phenomena such as
long-range order induced by order-by-disorder,~\cite{champion,savary_ObD}
multiple-$k$ long-range ordered phase,~\cite{stewart} 
 spin liquid behavior~\cite{gardner_TbTO_PRL}
and spin ice physics,~\cite{harris_SI,bramwell_PRL2001,ramirez_Nature} the latter  having attracted much 
interest.~\cite{gingras_springer,gardner_RMP,bramwell_Science,castelnovo_rev-cmp}
The spin ice  state displays a residual low-temperature magnetic entropy close
to that found for common water ice,~\cite{ramirez_Nature,cornelius,wiebe_prl} 
hence the name spin ice.
Most recently, a renewed flurry of theoretical and experimental  efforts have been directed at the study of
spin ices for they have been argued to display an emergent Coulomb phase~\cite{henley_ann_rev_cmp}
accompanied at low energies by deconfined fractionalized magnetic charge excitations, 
or ``monopoles".~\cite{castelnovo_rev-cmp,henley_ann_rev_cmp,castelnovo_nature}

Notwithstanding the rich physics that RE$_2$M$_2$O$_7$ materials display,~\cite{gardner_RMP} 
 they have, until very recently,~\cite{singh_perspective,ross_PRX,savary_PRL,lee_onoda}
not attracted that much interest from the community of theorists and 
experimentalists searching for quantum spin liquids.~\cite{balents}
Quantum fluctuations are expected to be, relatively speaking, 
more significant the smaller the spin quantum number ${\bm S}$. On the other hand, because of the intrinsic large
spin-orbit coupling at play in the Lanthanide 4f series, the  total angular momentum 
${\bm J} = {\bm L} + {\bm S}$, and not ${\bm S}$, is a good quantum number with $J$
being typically  large across the whole Lanthanide series 
(e.g. $J=8$  and $J=S=7/2$ for free Ho$^{3+}$ and Gd$^{3+}$ ions, respectively).
This situation would seem unpromising to those seeking magnetic materials with potentially
large quantum mechanical spin fluctuations and exotic quantum states of matter.
However, such a perspective is perhaps too pessimistic as we now discuss.

In insulating magnetic compounds with 4f elements, 
the various inter-ion interactions, $H_{\rm int}$ 
(exchange, superexchange, virtual phonon exchange and magnetostatic dipole-dipole coupling) 
are typically  weak compared to the single-ion crystal-field interactions defined by a crystal-field Hamiltonian $H_{\rm cf}$. 
As a first approximation, one thus often proceeds 
by determining the energy level spectrum of $H_{\rm cf}$ for a fixed $J$ manifold.~\cite{gingras_springer}
The point group symmetry of the crystal dictates the allowed symmetry properties of $H_{\rm cf}$.
These  symmetries determine, in return, the spectral decomposition of the crystal-field states that derive from the original 
$2J+1$ degenerate $^{2S+1}L_{J}$ electronic ground state of the otherwise free RE$^{3+}$ ion. 
In the simplest case, the resulting crystal field ground state is a magnetic doublet  with
wavefunctions $\vert \psi^{+}\rangle$ 
and $\vert \psi^{-}\rangle$ that have $\vert \psi^{\pm}\rangle = \sum_{m_J} C^{\pm}_{m_J} \vert J,m_J\rangle$ for spectral decomposition.
 $\vert \psi^{\pm}\rangle$  contains
{\it all} the $\vert J,m_J\rangle$ spectral components that transform similarly according to the point group symmetry operations.
Consequently,  there may or not be nonzero 
$\langle \psi^\pm \vert H_{\rm int} \vert \psi^\mp \rangle$  matrix  elements.
This depends on the specific nature of the inter-ion couplings $H_{\rm int}({J_i^u})$,  
a function of the components $J_i^u$ ($u=x, y, z$)
of angular momentum ${\bm J}_i$ of ion $i$,~\cite{what_pm_and_z_means} 
as well as the
specific ion-dependent
spectral decomposition of $\vert \psi^{\pm}\rangle$.
It is the nonzero
$\langle \psi^\pm \vert H_{\rm int} \vert \psi^\mp \rangle$
matrix elements~\cite{g-tensor} which determine whether significant quantum dynamics exist 
within the low-energy sector.
Most importantly, quantum dynamics need not be ruled out despite the large ${\bm J}$ of the isolated RE$^{3+}$ since the
crystal field Hamiltonian $H_{\rm cf}$ entangles a superposition of the $\vert m_J\rangle$ 
eigenstates of $J^z$.~\cite{what_pm_and_z_means}
  As a result, $H_{\rm int}$, by virtue of its lack of commutation with $H_{\rm cf}$, can, in principle,
have nonzero matrix elements between 
$\vert \psi^{+}\rangle$ and $\vert \psi^{-}\rangle$ and induce quantum dynamics within the 
low-energy Hilbert space
spanned by
 $\prod_i^N {\vert {\psi_i^+} \rangle} {\vert {\psi_i^-} \rangle}$.

In 4f ions with an even number of electrons (i.e. non-Kramers ion) such as Pr, Tb and Ho, time-reversal
symmetry imposes that $\langle \psi^\pm \vert J^\pm \vert \psi^\mp \rangle=0$.~\cite{g-tensor}
Consequently, in presence of solely bilinear interactions of
the form $K_{ij}^{uv}(r_{ij}) J_i^u J_j^v$ with anisotropic $K_{ij}^{uv}$ couplings,
 non-Kramers ions would display no quantum dynamics at low-energy and behave as effective
classical Ising spins S=1/2,  as in the well-studied LiHoF$_4$ dipolar Ising system.~\cite{LiHoF4_review}
In the (Pr,Tb,Ho)$_2$(Ti,Sn,Zr)$_2$O$_7$ materials,~\cite{no_Pr2Ti2O7} 
interactions beyond bilinear ones or consideration of the excited crystal
field states are necessary to cause quantum dynamics
in the low-energy sector.~\cite{molavian,mcclarty,onoda}
In that context, multipolar interactions  in Pr$_2$(Sn,Zr,Ir)$_2$O$_7$ compounds~\cite{onoda}
and virtual crystal field excitations in the Tb$_2$(Ti,Sn)$_2$O$_7$ materials
have been discussed.~\cite{molavian,mcclarty}
Conversely, odd-electron (Kramers) ion systems (e.g. Gd, Dy, Er, Yb) are in principle symmetry-allowed to have nonzero
$\langle \psi^\pm \vert J^\pm \vert \psi^\mp \rangle$ matrix elements. The famous Dy$_2$Ti$_2$O$_7$ spin ice compound,
in which the Dy$^{3+}$ ions are Kramers ions and for which
 the excited crystal field states lie at $\sim 300$ K above the
ground doublet, 
has been shown to be well-described by a dipolar spin ice model \cite{denHertog,yavorskii}  
with classical Ising spins.~\cite{gingras_springer}
This success very likely signals rather negligible interactions among the $J_i^u$ components
beyond bilinear ones, concomitantly with the
specific spectral decomposition of $\vert \psi^{\pm}\rangle$ for Dy$^{3+}$ in Dy$_2$Ti$_2$O$_7$.~\cite{rosenkranz,malkin,bertin}
 On the other hand, Er$_2$Ti$_2$O$_7$ and Yb$_2$Ti$_2$O$_7$ have been known for some time \cite{champion,malkin,bertin}
 to have predominant ``transverse'' 
$\langle \psi^\pm \vert J^\pm \vert \psi^\mp \rangle$ matrix elements,~\cite{g-tensor} 
along with 
a non-negligible $\langle \psi^\pm \vert J^z \vert \psi^\pm \rangle$ ``longitudinal'' ($g_{zz}$ tensor) component.
Yet, the two compounds display quite different behaviors.
 Er$_2$Ti$_2$O$_7$ has overall antiferromagnetic interactions and develops long-range order at 1.2 K 
with zero propagation
vector ${\bm q}_{\rm ord}$ and zero magnetic moment per tetrahedron~\cite{champion} 
that is induced by order-by-disorder.~\cite{savary_ObD}
In contrast, Yb$_2$Ti$_2$O$_7$ has overall ferromagnetic interactions and exhibits a phase transition at
$T_c \sim 0.24$ K.~\cite{hodges,blote}
However, the nature of the long-range order below $T_c$ remains 
disputed~\cite{yasui,gardner_YbTO,chang} and 
the high-sensitivity of the properties in the low-temperature regime ($T \lesssim 300$ mK) 
on sample quality is just beginning to be understood.~\cite{yaouanc_YbTO_sample,ross_YbTO_sample,ross_YbTO_stuff}
It is here, within the Yb$_2$M$_2$O$_7$  family 
with overall ferromagnetic interactions and significant 
$\langle \psi^\pm \vert J^\pm \vert \psi^\mp \rangle$ 
matrix elements,~\cite{ross_PRX,g-tensor} 
that the potential for  an exotic class of quantum spin liquid arises  -- a possibility 
that  may have been casually dismissed from the naive perspective
of ``there should be negligible quantum effects in such large ${\bm J}$ systems''.

\subsection*{Quantum spin ice and Yb$_2$Ti$_2$O$_7$}

The Dy$_2$(Ti,Sn,Ge)$_2$O$_7$ and Ho$_2$(Ti,Sn,Ge)$_2$O$_7$ materials are classical Ising systems which 
may be viewed in their spin ice regime as collective paramagnets \cite{villain} or, employing a 
more contemporary terminology,  classical spin liquids.~\cite{balents} 
As discussed above, highly frustrated magnetic systems, 
by virtue of their low-propensity to develop classical long-range order, are
attractive candidates to search for quantum spin liquid behavior. 
Spin ice, reinterpreted as a classical spin liquid,~\cite{balents}
 may thus be viewed as a natural setting to explore how the addition
of quantum dynamics may give rise to a quantum spin liquid state.
Such a topic was originally explored 
by Hermele, Fisher and Balents \cite{hermele} and Castro-Neto, Pujol and Fradkin \cite{castro-neto}
a few years ago in the context of minimal theoretical toy models.
These two groups argued in their respective paper that the addition of quantum dynamics within a parent (constrained)
classical spin ice manifold may promote the system to a U(1) quantum spin liquid.
Such a state would be describable
by a quantum field theory analogous to that of quantum electrodynamics (QED) in 3+1 dimensions.
As a consequence, this $U(1)$ spin liquid state would display 
``electric'' and ``magnetic'' charge excitations and an accompanying gauge boson, or ``artificial photon''.~\cite{hermele,castro-neto}
Numerical simulations have, over the past few years, suggested that such phenomenology may indeed
be at play in various  lattice toy models.~\cite{banerjee,shannon_QI}
Such a U(1) quantum spin liquid state, which may be 
referred to as ``quantum spin ice'',~\cite{molavian} has been
suggested to explain some of the properties of real materials such as Tb$_2$Ti$_2$O$_7$ \cite{molavian,mcclarty} 
and Pr$_2$(Sn,Zr)$_2$O$_7$.~\cite{onoda} 
Most recently, it has been proposed that Yb-based Yb$_2$M$_2$O$_7$ pyrochlore oxides \cite{ross_PRX,savary_PRL} 
may offer an exquisite class of systems to investigate the possible existence of a quantum spin ice state 
and where the complexities of virtual crystal field fluctuations~\cite{molavian,mcclarty}
and magnetoelastic coupling~\cite{bonville_tetrag,petit_tetrag} that complicate
 the Tb$_2$(Ti,Sn)$_2$O$_7$ compounds  are avoided.
In a theory paper building on the original work of Hermele {\it et al.},~\cite{hermele} 
Savary and Balents have recently put forward a mean-field lattice gauge 
theory which identifies a number of possible phases, the most exotic ones
being the aforementioned U(1) spin liquid as well as
 a novel Coulomb ferromagnetic state.~\cite{savary_PRL}
This approach has been further extended to non-Kramers ion systems.~\cite{lee_onoda}
Finally, the question of how to probe the emergent photon in quantum spin ice via 
inelastic neutron scattering measurements has very recently been discussed.~\cite{benton}

One particularly interesting aspect of the search for quantum spin liquids in the Yb-based pyrochlores, and perhaps
Pr-based pyrochlores as well,~\cite{onoda,lee_onoda,gap_case-Yb_Pr} 
is that the microscopic Hamiltonian is parametrized by a handful 
 of  independent
anisotropic exchange  parameters $\{J_e\}$
 (four for Yb$^{3+}$ and three for Pr$^{3+}$)
~\cite{ross_PRX,thompson_PRL,curnoe_PRB,mcclarty_ETO}
between effective spin-1/2 degrees of freedom on each pyrochlore lattice site.
Furthermore, it may be that the long-range dipolar interactions, so important in classical 
Dy$_2$(Ti,Sn)$_2$O$_7$ and Ho$_2$(Ti,Sn)$_2$O$_7$ spin ices with the large magnetic Dy$^{3+}$ and Ho$^{3+}$ magnetic moment,~\cite{gingras_springer,bramwell_PRL2001,denHertog,yavorskii,gingras_CJP,isakov_PRL}
can perhaps be neglected as a first approximation.
To make definite progress at this time, a determination of the effective anisotropic exchange from 
experiments and theory using controlled approximations 
is required in order to position a candidate spin ice material in the phase diagram of 
Ref.~[\onlinecite{savary_PRL}],
or the corresponding phase diagram relevant to non-Kramers ions.~\cite{lee_onoda}
In the case of Yb$_2$Ti$_2$O$_7$,  the determination 
of those interactions from a series of experiments \cite{cao_PRL,cao_JPCM,thompson_PRL,ross_PRX,chang} 
has led to different   $\{J_e\}$ sets with very different numerical values.
Perhaps most noteworthy, a determination of those parameters from a fit to spin waves in strong magnetic field measured
via inelastic neutron scattering \cite{ross_PRX}
 give values significantly different than those obtained by fitting the zero-field diffuse 
paramagnetic neutron scattering using a random phase approximation (RPA)
method~\cite{thompson_PRL} at a temperature $T\sim 1.4$ K 
or a subsequent polarized neutron scattering version, also using RPA
as the fitting procedure, but now very near $T_c$.~\cite{chang}
Encouragingly, however, recent numerical linked-cluster (NLC) calculations \cite{applegate}
have found  the zero-field magnetic specific heat data of Yb$_2$Ti$_2$O$_7$ 
to be well described above 0.7 K 
by the $\{J_e\}$ set obtained from the inelastic neutron scattering,~\cite{ross_PRX} 
but not by the other sets determined from RPA fits to diffuse neutron scattering.~\cite{chang,thompson_PRL}


It is clearly desirable to fit bulk measurements to determine the $\{J_e\}$ parameters 
in cases where inelastic neutron scattering studies are impractical, and
to corroborate such neutron studies and understand thermodynamic and bulk magnetic properties in a common framework with neutron studies
whenever possible.
Yet, the scarcity of controlled numerical methods readily available
to calculate the thermodynamic properties of frustrated 
three-dimensional anisotropic quantum spin systems in a temperature regime where non-trivial correlations develop
do not make this task straightforward.
With the seeming success of previous NLC calculations applied to Yb$_2$Ti$_2$O$_7$,~\cite{applegate} we are thus naturally led to
ask: 

\begin{narrowtext}
{\it Can one convincingly demonstrate that NLC does provide a controlled method to describe bulk data for
a three-dimensional frustrated quantum (spin ice) system as it progressively
enters its low-temperature strongly correlated regime?}
\end{narrowtext}

In this paper, we address this question by 
extending the work of Ref.~[\onlinecite{applegate}]
by computing the thermodynamic properties of 
Yb$_2$Ti$_2$O$_7$ in nonzero magnetic field using NLC and by comparing the results of such
calculations with measurements above 2 K on single crystals. 
We show, through a comparison with NLC, 
 that the effective Hamiltonian with its anisotropic exchange parameters $\{J_e \}$
previously determined by fitting spin waves in the field polarized paramagnetic state~\cite{ross_PRX}
 describes, with no adjustment, the temperature $T$ and magnetic field $h$ dependence of the
 magnetization, ${\bm M}(T,h)$, of Yb$_2$Ti$_2$O$_7$. As a consequence, we demonstrate simultaneously the usefulness
of the NLC method and further validate~\cite{applegate} the quantitative merit of the effective spin Hamiltonian of
Ref.~[\onlinecite{ross_PRX}] to describe Yb$_2$Ti$_2$O$_7$.


The rest of the paper is organized as follows. In the next section, we describe the effective spin-1/2 model
for Yb$_2$Ti$_2$O$_7$ along with the NLC method. In Section \ref{experimental}, we discuss the details of
the experimental method employed. The NLC results are presented in Section \ref{results} while
Section \ref{confront} provides a comparison between experiment and theory. 
 A brief discussion in Section \ref{discussion} concludes the paper.


\section{Model \& Computational Method}
\label{model}

Symmetry considerations imply that nearest-neighbor bilinear exchange constants on the pyrochlore lattice
can be parametrized in terms
of $4$ distinct exchange parameters and two $g$-factors. In the local basis,
this model is defined by the Hamiltonian \cite{ross_PRX,savary_PRL,gap_case-Yb_Pr,thompson_PRL,curnoe_PRB,mcclarty_ETO}
\begin{eqnarray}
{\cal H}_{\rm QSI}
&=&\sum_{\langle i,j\rangle} \{ J_{zz}S_i^z S_j^z - J_\pm (S_i^+ S_j^- +S_i^-S_j^+)\nonumber \\
        & &+ J_{\pm\pm} [\gamma_{ij}S_i^+S_j^+ + \gamma_{ij}^* S_i^-S_j^-]\nonumber \\
        & &+ J_{z\pm} [(S_i^z(\zeta_{ij}S_j^+ +\zeta_{i,j}^*S_j^-) + i \leftrightarrow j] \} .
\label{Hqsi}
\end{eqnarray}
Several notation conventions are possible for ${\cal H}_{\rm QSI}$.~\cite{ross_PRX,thompson_PRL,mcclarty_ETO,curnoe_PRB}
 Here, we 
adopt the one used  in Ref.~[\onlinecite{ross_PRX}].
In Eq.~(\ref{Hqsi}), $\langle i,j \rangle $ refers to nearest-neighbor sites of the pyrochlore lattice,
$\gamma_{ij}$  is a $4 \times 4$ complex unimodular matrix, and $\zeta=-\gamma^*$.~\cite{ross_PRX,savary_PRL}
The $\hat z$ quantization axis is along the local $[111]$ direction,~\cite{what_pm_and_z_means}
and $\pm$ refers to the two orthogonal local directions. The $g$ tensor takes values $g_{zz}$ along
the local $[111]$ cubic direction and $g_{xy}$ perpendicular to it.
In the presence of an applied external magnetic field ${\bm h}$, an additional Zeeman interaction 
$H_{\rm Z}=- {\bm h} \cdot {\overleftrightarrow{g}} \cdot {\bm S} \mu_{\rm B}$ is added to ${\cal H}_{\rm QSI}$, giving a total
spin Hamiltonian ${\cal H}={\cal H}_{\rm QSI} + {\cal H}_{\rm Z}$.
Here $\overleftrightarrow {g}$ is the $g$-tensor with eigenvalues 
$(g_{xy},g_{xy},g_{zz})$.~\cite{what_pm_and_z_means,g-tensor}

As already stated above,
there have been several experimental attempts to determine these  $\{ J_{\rm e} \}$ effective exchange
parameters leading to widely different results.~\cite{ross_PRX,chang,thompson_PRL,cao_PRL,cao_JPCM,malkin}
Among them, Ross {\it et al.} \cite{ross_PRX} used inelastic neutron scattering (INS) data 
obtained from measurements in high magnetic fields 
(i.e. in the polarized paramagnetic state) to determine the 
 $\{ J_{\rm e} \}$ couplings for Yb$_2$Ti$_2$O$_7$:
 $J_{zz}=0.17\pm 0.04$, $J_\pm=0.05\pm 0.01$, $J_{\pm\pm}=0.05\pm 0.01$,
and $J_{z\pm}=-0.14\pm 0.01$, all in meV and with $g_{zz}=1.80$ and $g_{xy}=4.32$. 
In a recent study,~\cite{applegate} it was found that the zero-field specific heat and entropy 
deduced from the data of Bl\"ote {\it et al.} \cite{blote}
are  well described only by the exchange
parameters obtained by Ross {\it et al.}~\cite{ross_PRX} 
A key goal of the present paper is to investigate further the validity of the Hamiltonian (\ref{Hqsi})
as a description of the bulk properties of 
Yb$_2$Ti$_2$O$_7$ with the $\{J_e\}$ exchange couplings from Ref.~[\onlinecite{ross_PRX}].
We do so by performing calculations of the
thermodynamic properties of model (1) as a function of temperature $T$ and magnetic field $h$
and compare the numerical results with those obtained from experimental measurements.
We find an excellent agreement without any
adjustment of the parameters determined by Ross {\it et al.},~\cite{ross_PRX}
hence providing a strong validation to the $\{J_e \}$ parameters determined by INS.
As a corollary, our work  confirms that fitting the INS spectra at high magnetic field is an excellent
method to determine the exchange constants for pyrochlore oxides with well isolated magnetic ground doublets
and which are well described by an effective spin-half model.~\cite{gap_case-Yb_Pr}
Further implications of this agreement will be discussed in the concluding Section \ref{discussion}.

To calculate the thermodynamic properties of model (\ref{Hqsi}), we turn to Numerical Linked-Cluster (NLC) 
expansions.~\cite{rigol} 
In this method, an extensive property $P$ (such as heat capacity or magnetization)
of a thermodynamic system is calculated as a sum over contributions
from different clusters embedded in the lattice:
\begin{equation}
P/N=\sum_c L(c) \times W(c).
\end{equation}
Here $L(c)$ is the count of the cluster per lattice site, defined as the number of
ways to embed the cluster.
$W(c)$ is the weight of the cluster which is obtained by calculating 
the property for a given cluster and subtracting the weight of all its subclusters.
\begin{equation}
W(c)= P(c) -\sum_ s W(s).
\end{equation}
Here, the sum runs over all subclusters of the cluster $c$. A subcluster
is defined as any cluster smaller than $c$ that can be embedded in cluster $c$. 
This scheme can be used
to develop power series expansions, such as high temperature series expansions in powers
of inverse temperature $\beta$, or, some coupling constant expansion (such as expansion in
inverse field-strength). It can also be
used to numerically calculate properties for a given value of temperature and coupling constants,
as we do in the present work.

The pyrochlore lattice is a lattice of corner-sharing tetrahedra
and it proves useful to develop
NLC in terms of clusters consisting of complete tetrahedra. We have calculated temperature
and field-dependent properties up to $4^{\rm th}$ order, that is, including contributions from all
clusters made of up to $4$ tetrahedra. These NLC calculations are numerically exact in two limits. 
They are so at high temperatures, since corrections to $4^{\rm th}$ order NLC is of order $\beta^6$ in the
high temperature series expansion for $\ln{Z}$. The NLC calculations are also exact at high magnetic field $h$ at
all temperatures since corrections to $4^{\rm th}$ order NLC 
for $\ln{Z}$ is of order $(J/h)^5$ at $T=0$ with
exponentially small corrections $\exp(-c h/T)$ at finite temperatures 
($c$ is some $h$ and $T$ independent  constant).
The parameter region where NLC begins to lose accuracy is when the temperature 
 and the applied field (Zeeman energy) are both smaller than the exchange constants. 
Thus, as long as either the high temperature expansion or the high-field expansion converges, 
the NLC calculations should be highly accurate. 

The reason for developing NLC in terms of complete tetrahedra comes from the fact that,
for spin ice systems, the `ice rule' constraints 
at the origin of spin ice physics are local to tetrahedra. Thus, having clusters with only
parts of a tetrahedron would cause wild oscillations in the calculated properties as the constraints
cannot be satisfied for such a cluster. Having only clusters with full tetrahedra allows the system to always satisfy
the ice-rule constraint. 
For example, in the case of the classical nearest-neighbor spin-ice exchange model,
such a tetrahedra-based NLC was found to be highly accurate.~\cite{oitmaa} In fact, the first order NLC,
which uses a single tetrahedron and is equivalent to Pauling's approximation for the
entropy of ice, gives quantities at all temperatures (in zero external field)
that are accurate to a few percent.~\cite{oitmaa}

However, in the quantum spin ice problem, NLC must break down at low temperatures due
to the development of either
long-range correlations or long-range entanglement. 
We have found (see results below and in Ref.~[\onlinecite{applegate}])
that the first signature of such a breakdown in the low temperature and low
field region is an alternation of the thermodynamic property $P$ considered. This reflects the fact
that, as system sizes increase, the thermodynamic properties must approach their infinite size
values in some way. Now, imagine that, at some order, the numerically calculated 
property is a bit too large, giving rise to
a large cluster weight $W¬©$. 
Subcluster subtraction  then causes the weight in the subsequent order to
be too small. This, in return, causes an alternation 
with the NLC order considered in the property obtained by restricting the sum to 
some order. Such an alternation is handled well by using the Euler summation method,~\cite{Euler_method} which ensures
that an alternating piece is completely eliminated from the partial sums at all orders. Thus, the Euler
resummed properties are only missing the longer range correlations which are necessarily absent
in finite order NLC. For Yb$_2$Ti$_2$O$_7$, we have found that in zero magnetic field~\cite{applegate} 
the thermodynamic properties converge down to 2 K without the Euler summation and down to 
about 1 K with Euler summation (the largest exchange constant for Yb$_2$Ti$_2$O$_7$,
$J_{zz}$ in Eq.~(\ref{Hqsi}), is approximately 2 K ~\cite{ross_PRX}). 

\subsection*{Computational details}

NLC using the tetrahedral basis requires exact diagonalization of increasingly
larger tetrahedral clusters.  Using a single
Intel\textsuperscript{\textregistered} Core\textsuperscript{\texttrademark} 
i7-920 processor and freely-available LAPACK linear algebra routines, 
diagonalizations for clusters of one tetrahedron (four sites) and two
tetrahedra (seven sites) could be done in less than a second, while the
three-tetrahedron (10-site) cluster still required less than 10 seconds.
Computing only the spectrum for a single four-tetrahedron (13-site) cluster
required approximately 25 minutes of cpu time and 2.1 GB of memory, or slightly over twice
the memory required to store the full Hamiltonian matrix ($2^{26}$ complex numbers).
Generating the full set of eigenstates required between 4 and 8 GB of memory.
With the computer resources available to us, full exact diagonalization of
larger systems, and thus higher orders of NLC expansion, were prohibited by memory
requirements.

\begin{figure} 
\begin{center} 
\includegraphics[width=8cm]{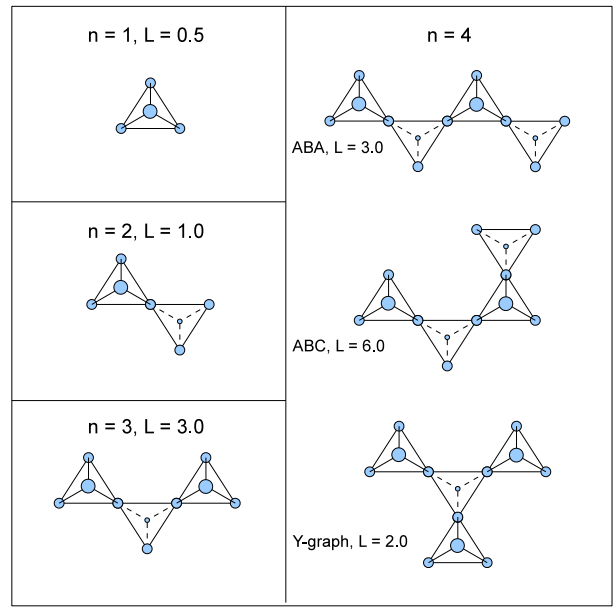}
\caption{\label{fig:clusters} (color online). 
List of topological clusters with complete tetrahedra.
In an applied field, each cluster must be treated as several distinct clusters
as explained in the text. The integer $n$ refers to the order at which the cluster
arises and $L$ gives the cluster count in zero-field.}
\end{center}
\end{figure}

A list of different topological clusters with complete tetrahedra is shown in Fig.~\ref{fig:clusters}.
A single site must be treated as a cluster as well, but is not shown.
Because the local quantization $\hat z$ axis varies from sublattice to sublattice
in the RE$_2$M$_2$O$_7$ pyrochlore oxides,~\cite{gardner_RMP,what_pm_and_z_means}
in the presence of a magnetic field, each cluster with different types of 
sublattice site must be treated separately. 
For example, each of the four single-site graphs (in NLC order $n=0$) must be treated separately
for a general field direction. A single tetrahedron graph remains unique even in a field.
But, the graph with two tetrahedra joined at a point (NLC order $n=2$), is really 4 distinct graphs that
depends on the type of site (i.e. sublattice) that is
 shared between the two tetrahedra. Similarly, all higher order
graphs must be split into several distinct graphs to complete the calculation.

\subsubsection*{Calculation of observables}
\newcommand{\tr}{\mbox{tr}}

We use the eigenvalue spectrum $\{E_{\alpha}\}$ of
${\cal H}={\cal H}_{\rm QSI} + {\cal H}_{\rm Z}$ 
to compute the partition function $Z$: 
\begin{equation} Z = \tr(e^{-\beta {\cal H}}) = \sum_{\alpha} e^{-\beta E_{\alpha} } . 
\label{eq:Z}
\end{equation}  
The expectation value for the internal
energy is computed from the formula 
\begin{equation} 
\left<E\right> =
\tr({\cal H} e^{-\beta {\cal H}}) / Z. 
\end{equation} 
 This quantity is used, in
turn, to compute the entropy as 
\begin{equation} 
\left<S\right>/k_{\rm B} = \beta \left<E\right> + \log Z, 
\end{equation} 
to which the heat capacity is related as 
\begin{equation} 
\left<C\right> = T \frac{\partial}{\partial T}\left<S\right>. 
\end{equation}

The average magnetization per site $M(T,h)$ is calculated
as a derivative of the free energy with respect to the applied magnetic field
$h$: 
\begin{equation} 
M(T,h) = \beta^{-1} \frac{\partial}{\partial h} \ln Z . 
\end{equation} 
 In practice, an approximation of this derivative at a
given field value requires two separate evaluations of the free energy
separated by a small change of field strength, in this case $10^{-6}$ T.

We believe in the correctness of the results from our calculations of
the field-dependent properties for the following three reasons.
First,  the same results were obtained from 
two independently written computer programs. Second, we checked that in
the limit $h\to 0$, the zero-field results~\cite{applegate} were reproduced. Third,  we checked
that at high temperatures the weights of the clusters were found to scale with 
the expected powers
 of inverse temperature $\beta$. These powers can be deduced from considerations of a high temperature expansion
of the quantity of interest. Consider, for example, an expansion for $\ln{Z}$.
In such an expansion either at least two powers of $\beta\equiv 1/(k_{\rm B} T)$ 
arise for each tetrahedron
or at least one power of $\beta$ arises for each tetrahedron together with
two powers of $\beta$ from placing the Zeeman ${\cal H}_{\rm Z}$ field-term 
on sites at the outside perimeter of the cluster,
as needed to give a non-zero contribution to the trace in Eq.~\ref{eq:Z}.
Thus, the weight of the 4-tetrahedra cluster is of order $\beta^8$ in zero field and
of order $\beta^6$ in non-zero field. These latter checks are non-trivial and demonstrate that
all subgraphs have been properly subtracted within the NLC procedure.

\subsubsection*{Euler summation}

NLC generates a sequence of property estimates $\{P_n\}$ with
increasing order $n$, where $P_n = \sum_{i=1}^n S_i$.  The convergence of such
a sequence can be improved by Euler summation.~\cite{Euler_method,applegate}
In general, given alternating terms $S_i = (-1)^i u_i$, the precise 
infinite-size lattice  property
$P_{\infty}(\mathcal{L})$ is approached by the sum (with $n$ even)
\begin{equation}
u_0 - u_1 + u_2 - \ldots - u_{n-1} + 
\sum_{s}^{\infty} \frac{(-1)^s}{2^{s+1}}[\Delta^s
u_n],
\end{equation}
where $\Delta$ is the forward difference operator
\begin{eqnarray}
 \Delta^0 u_n & = & u_n, \nonumber \\
 \Delta u_n & = & u_{n+1} - u_n, \nonumber \\
 \Delta^2 u_n & = & u_{n+2} - 2 u_{n+1} + u_n, \nonumber \\
 \Delta^3 u_n & = & u_{n+3} - 3 u_{n+2} + 3 u_{n+1} - u_n, \ldots .
\label{Euler_trans}
\end{eqnarray}
Usually, a small number of terms are computed directly, and the Euler
transformation, $P_{n,\text{E}}$ defined below, 
is applied to the  rest of the series.  In our case, where direct
terms were available up to fourth order, we began the transformation
after the second order,
so that the third and fourth order Euler-transformed property estimates, 
$P_{3,\text{E}}$ and $P_{4,\text{E}}$ respectively, are given by
\begin{eqnarray}
 P_{3,\text{E}} & = & S_0 + S_1 + S_2 + \frac{1}{2}S_3, \nonumber \\
 P_{4,\text{E}} & = & P_{3,\text{E}} + \frac{S_3+S_4}{4}.
\end{eqnarray}

\section{Experimental Methods}
\label{experimental}

As we are aiming to establish a close connection between experiment and theory for the temperature
and field dependence of the magnetization $M(T,h)$, it is necessary to perform adequate demagnetization
corrections and this, in turn, necessitates measurements on single crystals.
Magnetization experiments were thus carried out on three single crystal pieces.  
The samples were cut from two large single crystals grown by the optical 
floating zone method, using a growth procedure similar to that previously 
employed for growing single crystals of Tb$_2$Ti$_2$O$_7$.~\cite{crystal_growth}
One crystal, which provided samples aligned along [110] and [100], 
was grown at a rate of 6mm/h in oxygen pressure of 4atm.  
The second crystal, from with a piece aligned along [111] was cut, was grown at 5mm/h in 2atm of oxygen.  
The samples were aligned using X-ray diffraction to within 2 degrees of 
of each of the three high-symmetry directions;
[100], [110], and [111].
 Two single crystal samples were cut into rectangular prisms (``needles'') measuring 
0.74 mm $\times$ 0.74 mm $\times$ 3.74 mm and 0.69 mm $\times$ 0.65 mm  $\times$ 2.67 mm, respectively, 
with [111] and [100] directions oriented along their long axes.  The applied magnetic field was also oriented parallel to the long axis.  
The third single crystal piece was cut and polished into
a rounded triangular shape which we approximate, 
for the purposes of demagnetization corrections, as an ellipsoid with major 
axes $a$ = 1.85 mm, $b$ = 1.5 mm and $c$ = 0.8 mm. 
The $a$ direction made an angle of 22$^{\circ}$ with respect to both  [110] and the applied field direction within the $ab$ plane.  
The magnetization of the same [110] sample was also reported in Ross {\it et al.},\cite{ross_PRL} but without a demagnetization correction applied.  
The data was collected with a Quantum Design MPMS instrument,
which uses a SQUID magnetometer to measure the DC magnetization in magnetic fields up to 5 T and temperatures as low as 1.8 K.
The magnetization data for the [100] and [111] samples were corrected for
demagnetization effects  using an approximate formula  for the demagnetization field in rectangular prisms.~\cite{aharoni_1998}
The data for the [110] sample was corrected by approximating it as a very flat ellipsoid, 
with the field direction 22$^{\circ}$ from the (long) $a$ direction.
   In general, the susceptibility, $M(T,h)/h$, can be corrected to account 
for the demagnetization field using the following formula, in SI units,
\begin{equation}
1/ \chi  = 1/ \chi_A - N
\end{equation}
where $\chi$ is the actual susceptibility that we aim to compare with the results
of NLC calculations. 
$\chi_{A}$ is the apparent susceptibility (given by the measured $M/h_{\rm applied}$)
 and $N$ is the demagnetization factor, which depends on the sample geometry.~\cite{aharoni_1998}

\section{NLC results: thermodynamic properties}
\label{results}

\begin{figure}
\begin{center}
\includegraphics[width=8cm,height=6cm]{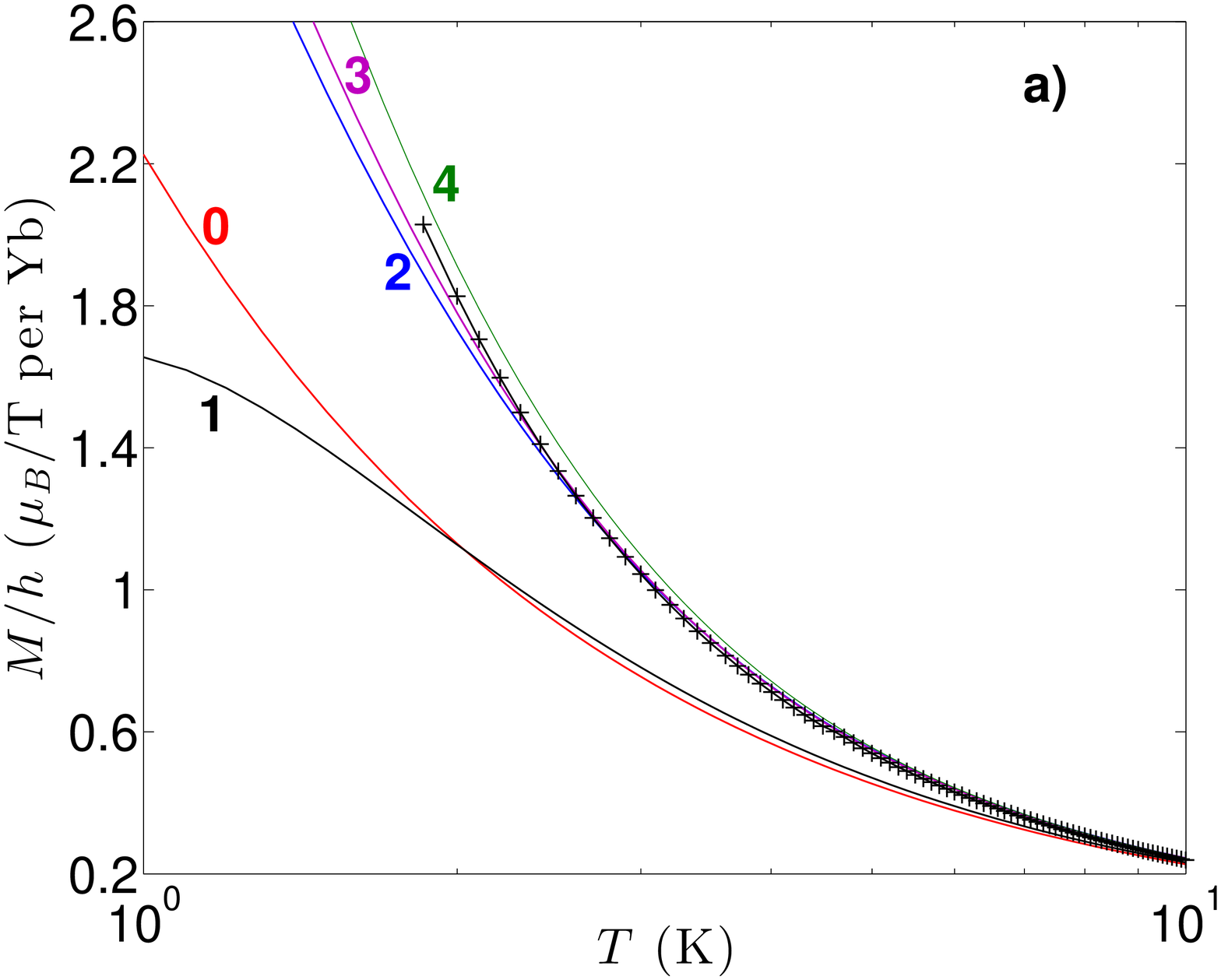}
\includegraphics[width=8cm,height=6cm]{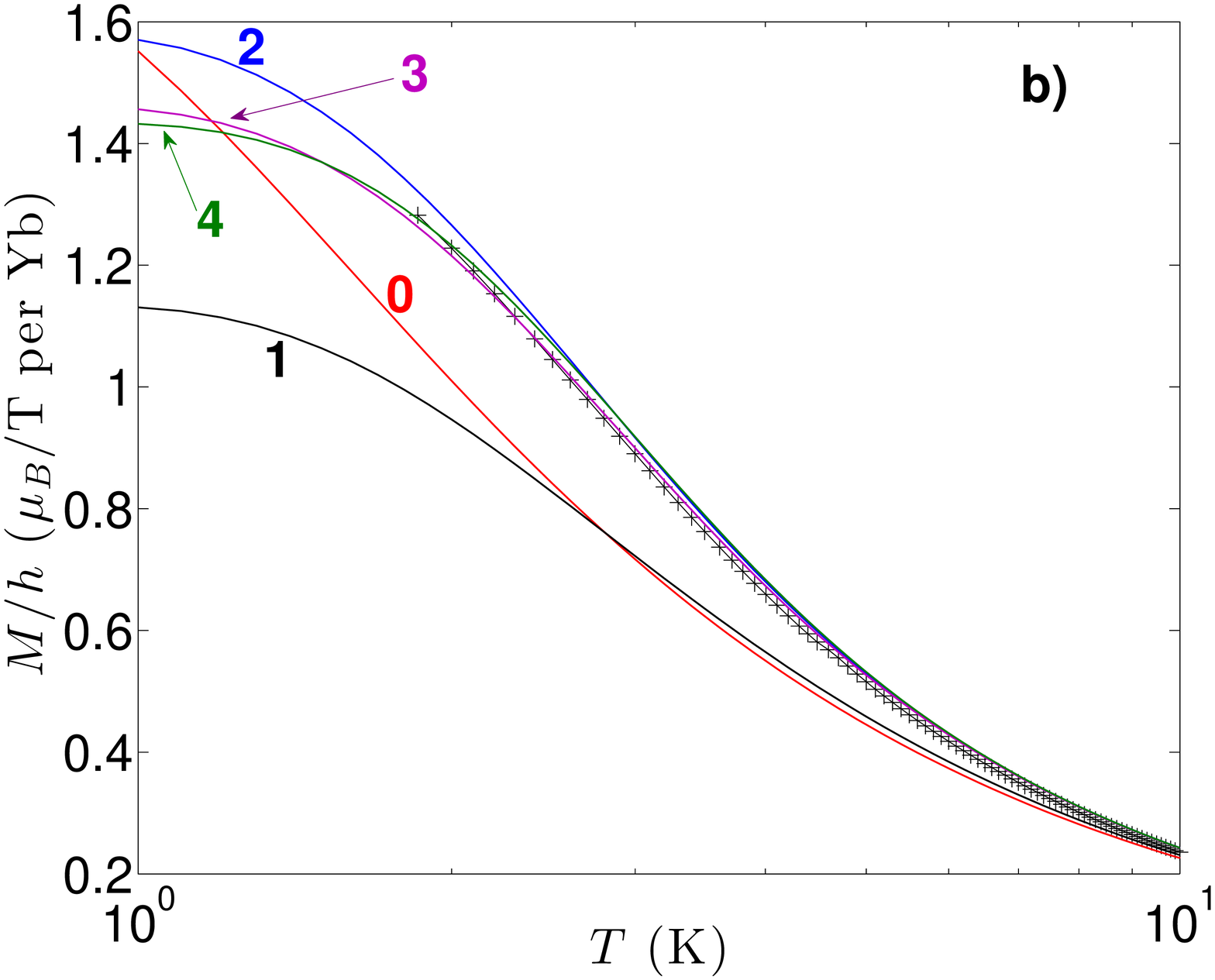}
\caption{\label{fig:merit_NLC} (color online).
The figure illustrates the evolution of the magnetization divided by the field strength,
 $M(T,h)/h$, with NLC order $n$  
and convergence towards experimental data. 
Panels a) and b) are for $h=0.2$ T and $h=1.0$ T, respectively, and with the field $h$ applied
along the $[100]$ direction.
Solid curves are for the NLC calculations and the number beside each curve corresponds
to the order $n$ up to which the calculations were carried to.
Pluses are the demagnetization corrected experimental $M(T,h)$ results divided by $h$.}
\end{center}
\end{figure}

The temperature dependence of the magnetization $M(T,h)$ divided by the strength of the applied magnetic field $h$
along the cubic $[100]$ direction and
calculated using the NLC method  is shown in Fig.~\ref{fig:merit_NLC}.
Panels a) and b) of Fig.~\ref{fig:merit_NLC} 
show the results for a field of 0.2 T and 1.0 T, respectively.
The experimental values of $M(T,h)/h$, after demagnetization corrections, are shown for the same
applied field values and are marked by black pluses.
The number beside each curve labels the NLC order at which $M(T,h)$ was calculated (see Section \ref{model}).
NLC-0 (red curve, label ``0'') considers a single site cluster.
It therefore does not incorporate the effect of the interactions in Eq.~\ref{Hqsi}
between the Yb$^{3+}$ ions and thus  
corresponds to the Yb$^{3+}$ single-ion property with $g$-tensor components 
$g_{zz}=1.80$ and $g_{xy}=4.32$.
The significant differences between NLC-0 and the experimental data emphasizes the importance of interactions
and concurrent development of correlations below 10 K.
The importance of interactions on causing a departure of 
$M(T,h)/h$ from its single-ion value below $T\lesssim 10$ K had previously been noted 
on the basis of a determination of the local susceptibility from 
polarized neutron measurements~\cite{cao_PRL,cao_JPCM} as well as 
a subsequent theoretical calculation of the local susceptibility \cite{thompson_JPCM} using 
exchange parameters determined from RPA fits to the diffuse paramagnetic scattering.~\cite{thompson_PRL}
However, none of these works~\cite{cao_PRL,cao_JPCM,thompson_JPCM} 
had the quantitative descriptive power of the present NLC calculations.

Compared to NLC-0, NLC-1 incorporates the effect of interactions and correlations at the scale of one 
tetrahedron (see Fig.~\ref{fig:clusters}). Focusing on the case $h=0.2$ T, one observes 
little difference between NLC-0 and NLC-1 above $T \sim 2$ K,  
but a large increase in $M(T,h)/h$ in going from NLC-1 to NLC-2 for $T\lesssim 7$ K.
The results for NLC-1 requires the exact diagonalization of the Hamiltonian over 
a single tetrahedron and therefore, incorporates spatial spin-spin correlations
extending only over a nearest-neighbor distance (see panel for $n=1$ in Fig.~\ref{fig:clusters}).
In contrast, NLC-2 considers two tetrahedra (see panel for $n=2$ in Fig.~\ref{fig:clusters}) and therefore 
considers correlations, and associated fluctuations, reaching out to third nearest neighbors.
As first noted by Ross {\it et al.},~\cite{ross_PRX} and further expanded upon by Applegate {\it et al.},~\cite{applegate} 
the fluctuations of the joining spin/site of two tetrahedra mediates an effective ferromagnetic third nearest-neighbor
coupling ($J_3$ in Ref.~[\onlinecite{applegate}]). 
This fluctuation-induced interaction promotes ferromagnetic correlations among the otherwise degenerate 
classical 2-in/2-out spin ice states, an effect that we believe important to induce
ferrimagnetic correlations in Yb$_2$Ti$_2$O$_7$.~\cite{ross_PRX,savary_PRL,applegate} 
The incorporation of this fluctuation process and induced effective $J_3$ coupling
only happens for NLC order $n \ge 2$ and is thus absent for NLC-1. 
 We believe that the large increase of $M(T,h)/h$ in going from NLC-1
to NLC-2 results from the ability of clusters of two-tetrahedra to support that type of fluctuation physics while
a single tetrahedron cannot. 
One may want to argue that the observed large increase of the calculated $M(T,h)/h$, when going from NLC-1
to NLC-2, is further evidence that Yb$_2$Ti$_2$O$_7$, as described by the effective exchange parameters of Ref.~[\onlinecite{ross_PRX}],
is either on its way or at the verge of developing spontaneous ferrimagnetic order at low-temperature.
We return to this point below in Subsection \ref{field:Cv} when we discuss the magnetic entropy, $S(T,h)$,
in nonzero magnetic field $h$.

Considering the results of Fig.~\ref{fig:merit_NLC}a), we observe that 
the NLC-2, 3 and 4 results are quite close to each other down to $T \sim 1.8$ K, 
the lowest temperature available for the experimental data. 
Most importantly, one should note that NLC-2, 3 and 4, using the effective anisotropic exchange
parameters $\{ J_e \}$ of Ref.~[\onlinecite{ross_PRX}] already provide, without Euler summation, 
and {\it without any} adjustment of
the parameters defining ${\cal H}_{\rm QSI}$ and ${\cal H}_{\rm Z}$, 
a very good agreement with the experimental data. 
 A similarly good agreement is also found between the experimental data 
and the NLC results for the highest NLC orders ($n=3, 4$) 
for a field $h=1$ T (see panel b) of Fig.~\ref{fig:merit_NLC}).

Having demonstrated that (i)  interactions do play a strong effect in renormalizing
$M(T,h)$ for $T\lesssim 10$ K and that (ii) NLC orders $n=3$ and $n=4$ give a highly suitable description
of the experimental data even for the weakest field $h=0.2$ T considered (see the discussion in Section \ref{model}),
we henceforth solely report results from the Euler transformation method 
(to order $n=3$ and $n=4$, see Eq.~(\ref{Euler_trans})) 
to generate theoretical values of $M(T,h)/h$ vs temperature 
and for various  field $h$ along the three high-symmetry cubic directions ($[100]$, $[110]$, and $[111]$).

\begin{figure} 
\begin{center} 
\includegraphics[width=8cm]{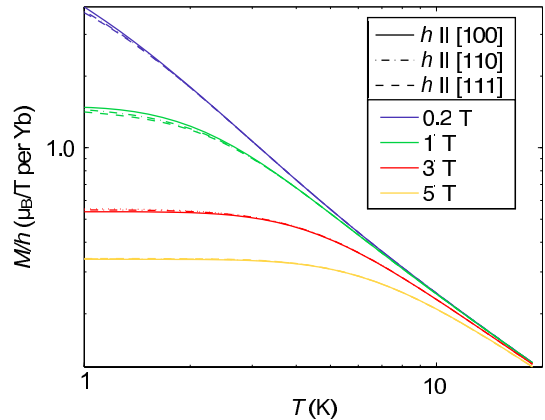} 
\caption{\label{fig:Euler_results} (color online).
Calculated magnetizations as a function of temperature for different field strengths and directions 
are shown using the 4$^{\rm th}$ order Euler transformed results from the NLC expansion (see. Eq.~(\ref{Euler_trans})).
  The magnetic fields are applied along the $[100]$, $[110]$, and $[111]$ directions.  }
\end{center}
\end{figure}

Figure \ref{fig:Euler_results}  shows $M(T,h)/h$, obtained using the $4^{\rm th}$ order
Euler transformed NLC results, as a function of temperature $T$ for fields $h$ 
of strength 0.2, 1, 3, and 5 T oriented parallel to the $[100]$, $[110]$ and $[111]$ directions.
 Again, the calculation employed the microscopic Hamiltonian 
appropriate to Yb$_2$Ti$_2$O$_7$ as derived from inelastic neutron scattering data at high field.~\cite{ross_PRX}
Figure \ref{fig:Euler_results} shows 4$^{\rm th}$ order NLC results for 
$M(T,h)/h$ per Yb$^{3+}$ ion vs temperature $T$, with both quantities plotted on a log scale.  
The calculated magnetization is relatively independent of direction and levels off at low temperatures, with the temperature 
at which it levels off being lower with smaller applied magnetic fields.  While the magnetization is only weakly dependent 
on the direction of the applied magnetic field, such differences in the magnetization 
as a function of direction of field are only evident at low temperatures.

\section{Magnetization: Confronting theory with experiment}
\label{confront}

\subsection*{Field dependent magnetization}

\begin{figure} 
\begin{center} 
\includegraphics[width=9cm]{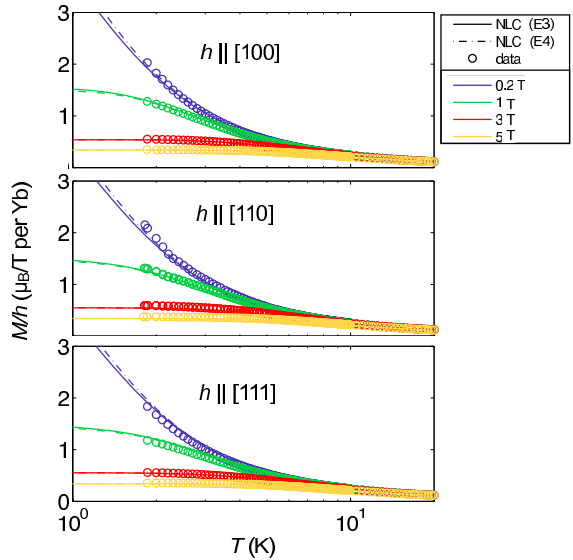}
\caption{\label{fig:3panels} (color online).
Calculated magnetizations using the $n=3$ and $n=4$
 Euler-transformed $M(T,h)/h$ NLC expansion results (see Eq.~(\ref{Euler_trans})) are  compared 
with the measured magnetization versus temperature for fields applied along the $[100]$, $[110]$ and $[111]$ directions.  
The parameters for ${\cal H}={\cal H}_{\rm QSI} + {\cal H}_{\rm Z}$ are those determined form the spin wave results of Ref.~[\onlinecite{ross_PRX}].
Appropriate demagnetization corrections have been applied to all experimental data.
Departures between $n=3$ and $n=4$ NLC expansion results become noticible at temperatures $\lesssim 1.8$ K.}
\end{center}
\end{figure}

Figure \ref{fig:3panels}
  shows the experimentally-determined magnetization as a function of temperature on a semi-log plot.  
Data is shown for applied magnetic field directions parallel to $[100]$, $[110]$, and $[111]$, 
from top to bottom, and for field strengths of 0.2, 1, 3, and 5 T.  Once again  $M(T,h)/h$  is normalized per Yb$^{3+}$ ion.
Two sets of NLC expansion calculations using the Euler transformation are shown in 
Fig.~\ref{fig:3panels}; to order $n=3$ are shown as the solid lines,
 and  to order $n=4$ are shown as the dot-dashed lines. 
 Here again,  the NLC expansions utilized the microscopic Hamiltonian derived from 
the high field inelastic neutron scattering (INS) data \cite{ross_PRX}
 with no adjustment in the parameters so determined.
 
 Several features are immediately clear from this comparison between theory and experiment.  
First, the (Euler-transformed) NLC expansion results to both $n=3$ and $n=4$ orders 
provide an excellent description of the magnetization 
for all field strengths and directions as well as all temperatures considered.
 The fact that a remarkable degree of quantitative agreement between theory and experiment
 is achieved without adjustment of the microscopic Hamiltonian appropriate to Yb$_2$Ti$_2$O$_7$ 
is strong validation of the determination of the microcopic 
Hamiltonian.~\cite{ross_PRX} 
 As the INS measurements were performed at $T=0.03$ K and applied magnetic fields of $h=$ 2 and 5 T with $h$ parallel to $[110]$,
 and the magnetization measurements are performed for T $>$ 1.8 K and fields $\le$ 5 T, we now have a very accurate description of Yb$_2$Ti$_2$O$_7$ using the same microscopic Hamiltonian over a remarkably large region of its $h-T$
phase diagram.

Looking at the Euler-transformed NLC expansion results for $n=3$ (solid line) and $n=4$ (dot-dashed line) in Fig.~\ref{fig:3panels}, 
one can see that, as expected, the two calculations are quite consistent with each other for  $T \gtrsim$ 1.8 K, but depart from each other at lower temperatures. 
 There is also a better agreement at lower temperatures between $n=3$ and $n=4$ NLC expansions for higher fields, 
independent of the direction of the applied magnetic field,
as anticipated on the basis of the general arguments presented in Section \ref{model}.

\subsection*{Other parametrizations of ${\cal H}_{\rm QSI}$}
\label{other_forms}

There have been 
over the last four years, apart from Ref.~[\onlinecite{ross_PRX}], 
a number of other studies~\cite{chang,thompson_PRL,cao_PRL,cao_JPCM,malkin} combining experiment and theory and which were aimed at determining the  strength of the anisotropic interactions in Yb$_2$Ti$_2$O$_7$. 
We believe that most, if not all, including Ref.~[\onlinecite{thompson_PRL}] that was co-authored by one of us, 
are beset by significant drawbacks compared to the in-field inelastic neutron scattering measurements of Ross {\it et al.}~\cite{ross_PRX}
In fact, one might have wondered whether the anisotropic exchange parameters $\{J_e\}$ extracted in Ref.~[\onlinecite{ross_PRX}] in magnetic
fields of 5 Tesla might have suffered from large renormalization due to magnetoelastic effects. 
The good agreement between experimental and NLC results for zero magnetic field specific heat~\cite{applegate} and the ones presented in this
paper for the temperature and magnetic field dependent magnetization, ${\bm M}(t,h)$, provide compelling evidence that such renormalization, if it
exists, is well within the experimental uncertainty of the estimated $\{J_e\}$ parameters.~\cite{ross_PRX}
That being said, we now discuss each of the other works.

Cao {\it et al.} used polarized neutron diffraction to extract the local susceptibility tensor of a number of 
RE$_2$Ti$_2$O$_7$ pyrochlores, including Yb$_2$Ti$_2$O$_7$,  in an applied external field of 1 Tesla along the
[100] crystallographic direction.~\cite{cao_PRL,cao_JPCM} 
Their analysis was based on a mean-field theory that ignores the sublattice nature of the pyrochlore lattice and incorporates
the effect of the local mean-field only via two independent coupling constants referred to as $\lambda_z$ and $\lambda_\perp$.
By construct, such an approach makes it difficult to reconstruct the microscopic exchange parameters of the spin Hamiltonian.
Furthermore, as they do not comment on this,
it is not clear that their data analysis took  into account demagnetization effects.
Malkin {\it et al.} reinvestigated the description of the bulk and local susceptibility of RE$_2$Ti$_2$O$_7$ compounds, but now starting from
a microscopic formulation, and incorporating an adequate sublattice structure in their model as well as including demagnetization corrections.~\cite{malkin} 
In the case of Yb$_2$Ti$_2$O$_7$, they reported being unable to 
describe the longitudinal site susceptibility $\chi$ of Cao {\it et al.} by using a single set of crystal field parameters.
It is perhaps important to note that Malkin {\it et al.} assumed that the bilinear anisotropic interactions between the magnetic moment operators were symmetric, 
which amounts to neglecting Dzyaloshinskii-Moriya (DM) -like  interactions.
As was found in the work of Ross {\it et al.},~\cite{ross_PRX} the $J_{z\pm}$ interactions, which originate 
from the DM interactions, are the second largest ones in ${\cal H}_{\rm QSI}$, and of almost the same magnitude as the largest $J_{zz}$ coupling.
Hence, the neglect of the antisymmetric interactions with coupling $J_{z \pm}$
for Yb$_2$Ti$_2$O$_7$ is bound to cause difficulty in obtaining a quantitative description
of the  properties of this material.
Consequently, because of the aforementioned caveats, the results obtained in 
Refs.~[\onlinecite{cao_PRL,cao_JPCM,malkin}] 
cannot not provide
 a quantitative account of the microscopic interactions at play in Yb$_2$Ti$_2$O$_7$.

Thompson {\it et al.}~\cite{thompson_PRL} were the first to 
consider a microscopic theory which incorporates the single-ion crystal field Hamiltonian, $H_{\rm cf}$, 
for Yb$^{3+}$ in Yb$_2$Ti$_2$O$_7$, the four symmetry-allowed bilinear nearest-neighbor interactions, 
$K_{ij}^{uv}$ between the ${\bm J}_i$ angular
momentum operators along with the long-range magnetostatic dipole-dipole interactions.~\cite{thompson_PRL}
In order to determine the $K_{ij}^{uv}$, Thompson {\it et al.} used a random phase approximation (RPA) to calculate the 
diffuse (energy-integrated) neutron scattering intensity $S({\bm q})$ in the $[hhl]$ scattering phase and compare with experimental measurements at 
a temperature $T=1.4$ K.~\cite{thompson_uw}
With  the Curie-Weiss temperature  of Yb$_2$Ti$_2$O$_7$ given by
$\theta_{\rm CW} \sim 0.5$  K $\pm$ 0.2 K, 
it would have seemed that RPA, which is in essence a mean-field scheme, may be a priori reasonably 
quantitatively accurate at a temperature of 1.4 K with a 
frustration/fluctuation level set by
$\theta_{\rm CW}/T \lesssim 0.3$. Forced by construction to describe short-range diffuse scattering, 
the RPA fit of Ref.~[\onlinecite{thompson_PRL}] provided a set of
couplings $K_{ij}^{uv}$ allowing for a good fit of $S({\bm q})$ and, by ``retroactive consistency'', 
a mean-field critical temperature $T_c \sim 1.1$ K.
When translating the determined $K_{ij}^{uv}$ parameters in the $\{ J_e \}$ 
notation of Ross {\it et al.}, one finds a large difference
between the two sets. The authors of Ref.~[\onlinecite{chang}] commented 
in the supplementary material section of their paper that a possible reason for the failure of the model of Thompson {\it et al.}~\cite{thompson_PRL} 
to agree with the results of Ross {\it et al.}~\cite{ross_PRX} is that the former work neglected multipolar 
interactions between the ${\bm J}_i$  operators beyond the considered bilinear ones. 
However, as discussed in the supplementary material of Ref.~[\onlinecite{thompson_PRL}], such critique~\cite{chang} 
would appear of little merit for the following reason.
Thanks to the essentially total isolation of the Kramers crystal field doublet of Yb$^{3+}$ from the
excited states, the ground crystal field doublet acts for an almost exact {\it invertible} unitary 
transformation of the $K_{ij}^{uv}J_{i,u}J_{j,v}$ interactions, hence providing 
a one-to-one correspondence between the bilinear $K_{ij}^{uv}$ couplings and the
$\{J_e\}$ effective anisotropic exchange between effective spin-1/2 operators of Ross {\it et al.}~\cite{ross_PRX}
In other words, the bilinear $K_{ij}^{uv}J_{i,u}J_{j,v}$ model of Ref.~[\onlinecite{thompson_PRL}] can be viewed as a 
``high energy''  model whose projection (with `correct' values of the  $K_{ij}^{uv}$ couplings) 
gives the $J_{e}S_i^u S_j^v$ interactions in 
the model of Eq.~(\ref{Hqsi})~[\onlinecite{ross_PRX}].

The difficulty with Thompson {\it et al.}'s results is readily understood on the basis of the mean-field theory results presented
in Ref.~[\onlinecite{ross_PRX}].
The mean-field $T_c^{\rm mf}$ using the $\{J_e\}$ parameters from Ref.~[\onlinecite{ross_PRX}]
 is approximately 3.5 K.
Such a high $T_c^{\rm mf}$ compared to the 1.4 K RPA fit of $S({\bm q})$ means that the 
extracted $K_{ij}^{uv}$ parameters suffer from 
a very significant and uncontrolled renormalization from thermal (and possibly quantum) fluctuations. 
Concerns that 1.4 K might be too low for a quantitative RPA fit of $S({\bm q})$ of Yb$_2$Ti$_2$O$_7$
 and that fits at the higher temperature of 9.1 K data~\cite{thompson_PRL} might
have been more appropriate had
been expressed in Ref.~[\onlinecite{thompson_uw}].
Unfortunately, the diffuse signal at 9.1 K proved too weak to proceed.
As demonstrated by Applegate {\it et al.} in their NLC calculations,~\cite{applegate} 
the anisotropic exchange parameters of Ref.~[\onlinecite{thompson_PRL}]
provide a poor description of the  zero-field magnetic specific heat $C(T)$ of Yb$_2$Ti$_2$O$_7$.

\begin{figure}
\begin{center}
\includegraphics[width=8cm]{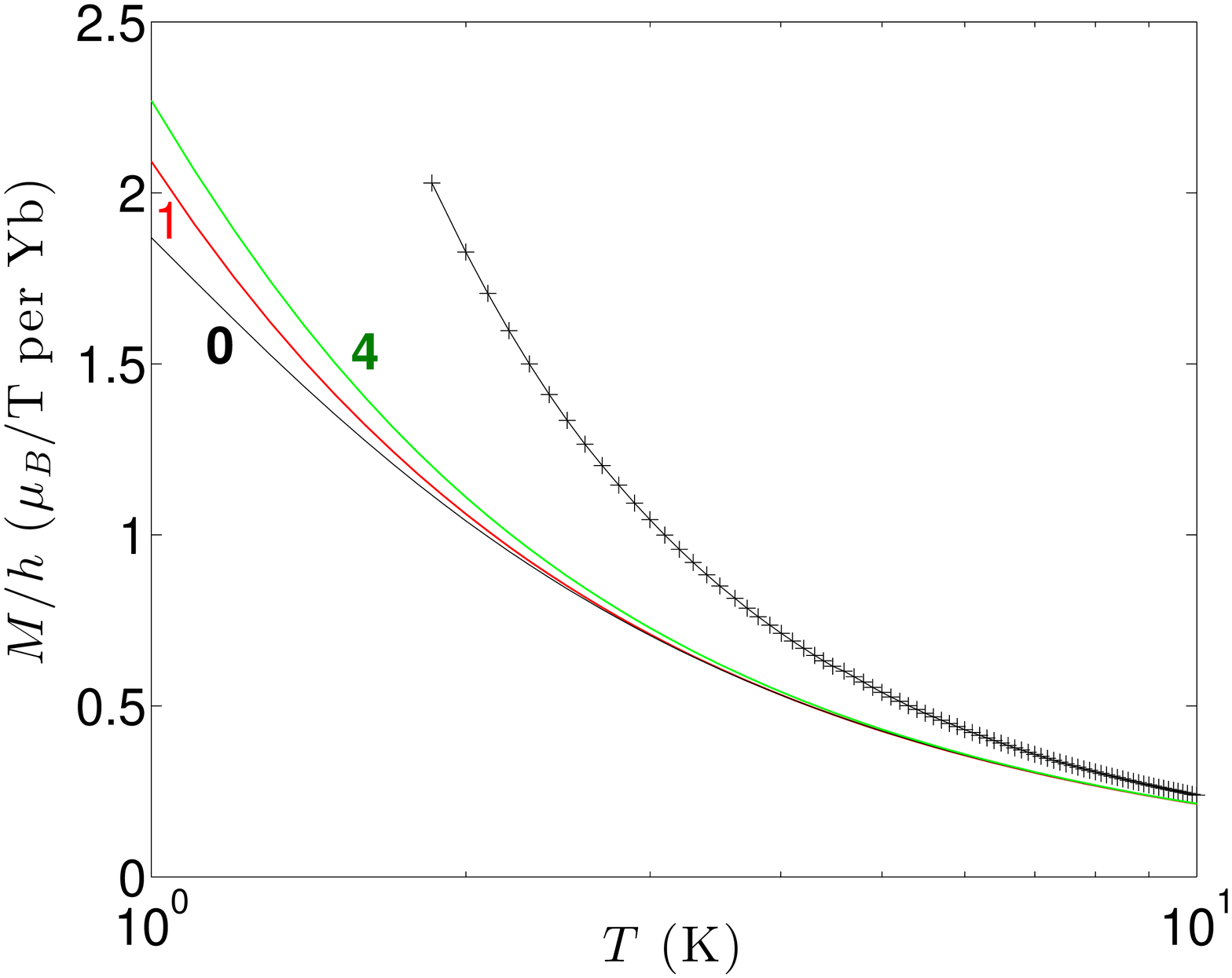}
\caption{\label{fig:chang_no_inter} (color online). The figure shows that $M(T,h)$ 
from NLC calculations (orders $n=0, 1, 4$) using the $\{J_e \}$ parameters reported by Chang {\it et al.}~\cite{chang} 
fail dramatically to describe the experimental $M(T,h)$ results (plus symbols). See text.}
\end{center}
\end{figure}

Chang {\it et al.} most recently reported their own estimate of the exchange parameters for  Yb$_2$Ti$_2$O$_7$.~\cite{chang}
They also employed an RPA scheme to fit the (polarized) neutron scattering intensity of the compound, but using an effective spin-1/2
model rather than the bilinear $K_{ij}^{uv}J_{i,u}J_{j,v}$ interactions plus $H_{\rm cf}$ of Thompson {\it et al.}~\cite{thompson_PRL} 
Perhaps believing in the incorrectness of the latter description (see discussion two paragraphs above), they passed over 
in the body of their paper the opportunity to offer a critique of Thompson {\it et al.}'s model. 
That said, with the published evidence from Ref.~[\onlinecite{ross_PRX}]
that $T_c^{\rm mf}$ for Yb$_2$Ti$_2$O$_7$ may be as high as 3.5 K~\cite{ross_PRX}, and that the RPA fits of Thompson {\it et al.} at 1.4 K may 
thus be of questionable quantitative merit, it is surprising that Chang {\it et al.} did nevertheless 
proceed to use RPA to fit the polarized neutron
scattering at a temperature as low as 0.3 K, which is a mere $25$ percent 
higher than the experimental $T_c\sim 0.24$ K.
It is also not clear how the authors of Ref.~[\onlinecite{chang}] find a similarity between their  $\{J_e\}$ values
and those of Ross {\it et al.}, especially after evidence had been reported by Applegate {\it et al.},~\cite{applegate}
that the $\{J_e\}$ of Ref.~[\onlinecite{chang}]  fail to describe $C(T)$,
while those of Ross {\it et al.} provide a quite adequate description of $C(T)$.~\cite{applegate}
Chang {\it et al.} propose a rationalization of the low-temperature state of Yb$_2$Ti$_2$O$_7$ 
on the basis of a Higgs-like phase and suggest positioning this material in a $\{J_e\}$ parameter space near a parent 
classical spin ice. Yet, at the same time, they overlook to comment on the inability of their 
microscopic model with its RPA-determined set of $\{J_e\}$ to describe $C(T)$.~\cite{applegate}
As in the case of Thompson {\it et al.}'s RPA analysis,\cite{already_critiqued} 
Chang {\it et al.}'s fit to the neutron scattering data
provides, by the very consequence of using RPA at a temperature $T \ll T_c^{\rm mf}$, for a set of coupling 
parameters which are significantly and uncontrollably renormalized downward compared to the bare $\{ J_e \}$.
This is illustrated in Fig.~\ref{fig:chang_no_inter}  where it is shown that NLC-1 and NLC-4 using the $\{J_e\}$
parameters of Chang {\it et al.} do not describe $M(T,h)$ for temperatures $T\lesssim 10$ K, where
correlations have barely started to develop.~\cite{thompson_PRL} 
In fact, down to $T=1$ K, there is little difference between either NLC-1 and NLC-4 with the 
single-ion magnetization with all interactions turned off and given by the NLC-0 results.

To conclude this discussion, it thus appears that only Ross {\it et al.}'s  set of microscopic parameters~\cite{ross_PRX} 
consistently describe Yb$_2$Ti$_2$O$_7$ for fields $h<5$ Tesla and temperatures $T\gtrsim 0.7$ K.

\subsection*{Field dependent specific heat}
\label{field:Cv}

\begin{figure}
\begin{center}
\includegraphics[width=9cm,angle=0]{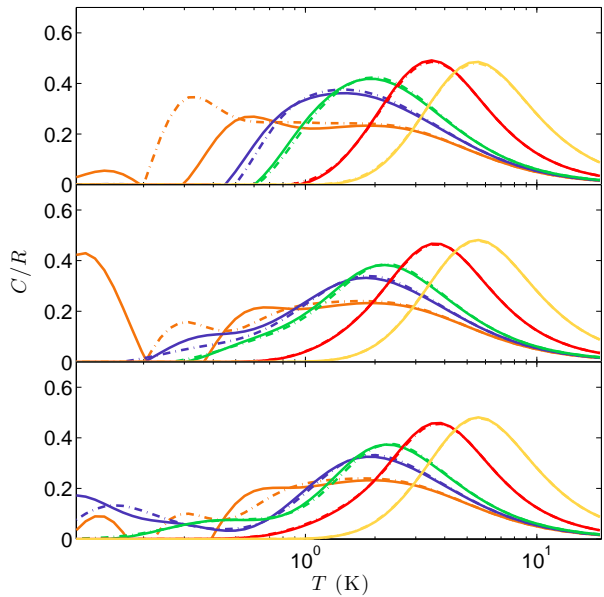}
\caption{\label{fig:spec_heat}
(color online).
The calculated molar specific heat using NLC is shown as a function of temperature and magnetic field
with field applied along, from top to bottom, the $[100]$, $[110]$ and $[111]$ directions. 
These results were calculated using the $n=3$ and $n=4$
 Euler transformation NLC results using the $\{J_e \}$ parameters of Ref.~[\onlinecite{ross_PRX}].
See the legends in top panel of Fig.~\ref{fig:entropy} for values of the field strengths
considered and legend for NLC order employed.}
\end{center}
\end{figure}

\begin{figure}
\begin{center}
\includegraphics[width=9cm,angle=0]{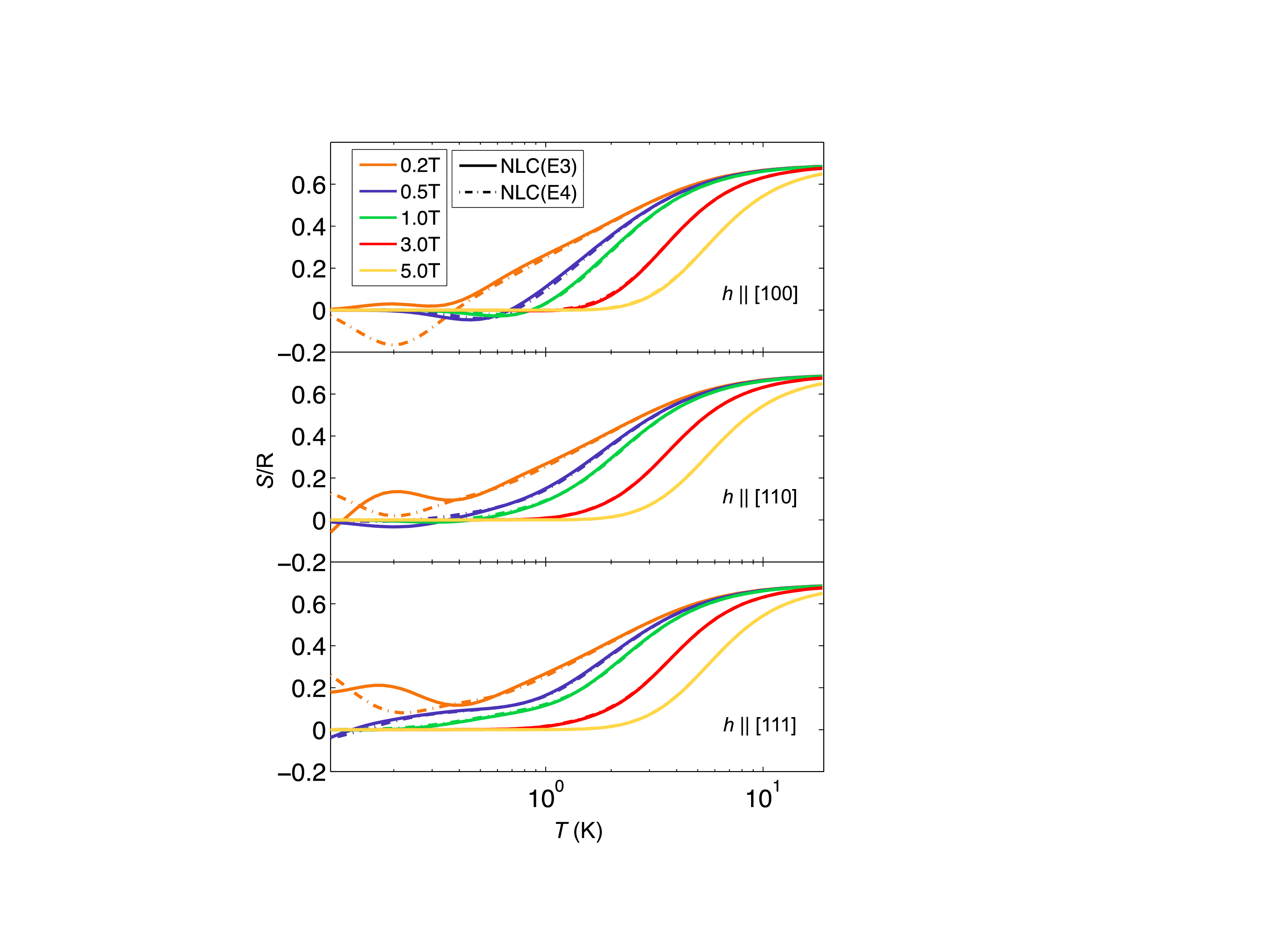}
\caption{\label{fig:entropy}
(color online).
The calculated molar entropy using NLC is shown as a function of temperature and magnetic field
with field applied along, from top to bottom, the $[100]$, $[110]$ and $[111]$ directions. 
These results were calculated using the $n=3$ and $n=4$ 
Euler transformation NLC results using the $\{J_e \}$ parameters of
Ref.~[\onlinecite{ross_PRX}].
}
\end{center}
\end{figure}

While we are not aware of in-field specific heat, $C(T,h)$, measurements on Yb$_2$Ti$_2$O$_7$ single crystals, we expect these
to be soon carried out given the interest devoted to this compound.
One perspective as to why such measurements might be of interest is the following.
The Er$_2$Ti$_2$O$_7$ pyrochlore antiferromagnet displays a transition to long-range order at $T_c \sim 1.2$ K.~\cite{champion,ruff,dalmas_ETO}
The application of a magnetic field along $[110]$ ultimately destroys that order at $T=0$, giving rise to a quantum phase
transition at $h_c \sim 1.5$ T.~\cite{ruff,dalmas_ETO} 
Specific heat measurements in non-zero field have been shown useful to characterize the evolution
of this system in and out of the long-range ordered phase at $h \lesssim 1.5$ T and $T \lesssim 
1.2$ K.~\cite{ruff,dalmas_ETO}

Measurements of $C(T,h)$ in Yb$_2$Ti$_2$O$_7$ for $T<0.3$ K may well be very interesting. 
This would be especially true if, once the sample dependence of $T_c$ has been understood and  
$T_c ~ 0.26$ K indeed turns out to be a phase transition to a ferrimagnetically-ordered state, 
as suggested by some experiments and theory, but not all experiments.~\cite{hodges,but_maybe_not}
Thus, in anticipation of such measurements in Yb$_2$Ti$_2$O$_7$, we have used NLC expansion to calculate  $C(T,h)$.

The specific heat, $C(T,h)/R$, calculated using the Euler-transformed 
NLC expansion to order $n=4$  and the microscopic Hamiltonian determined 
for Yb$_2$Ti$_2$O$_7$ from inelastic neutron scattering (INS) \cite {ross_PRX}
 is shown in Fig.~\ref{fig:spec_heat}. 
 We present results for $C(T,h)/R$ for a field parallel
to $[100]$, $[110]$, and $[111]$ from top to bottom, respectively.  
In all cases, calculations are presented for field strengths ranging from 0.2 T to 5 T.  

The trend is very similar for all three field directions.  
A broad peak is observed for a field strength $h$ above $\sim$ 2 T, with the peak position moving to lower 
temperatures as  $h$ is decreased.  
While the NLC calculation is known to be inaccurate for temperatures less than $\sim$ 1 K \cite{applegate} and for appropriately low field strengths, 
the calculated molar specific heat at the lowest applied magnetic fields is qualitatively consistent with that observed experimentally 
in zero field,~\cite{blote} displaying a sharp anomaly at very low temperaures as well as a broad shoulder near 2 K.

 Even though, at first sight, the dependence of
thermodynamic properties looks independent of field direction, there are subtle differences
at low temperatures and fields, which could be taken as evidence of an ultimate spontaneous ferrimagnetic 
ordering  with $M$ along [100] in zero field.~\cite{ross_PRX,savary_PRL,chang,applegate,but_maybe_not}
On close inspection, one finds that for fields of 1 T and lower, $C(T,h)$ is higher
at temperatures around 1 K for the [100] direction and, as a result, the entropy $S(T,h)$ is substantially
lower. In fact, as shown in Fig.~\ref{fig:entropy}, 
the entropy along [100] appears to become vanishingly small at temperatures
of order 1 K, whereas there remains a residual entropy along the other two directions.
The residual entropy remains largest along the [111] direction. 
This observation is consistent with the picture of a zero-field ordering in this system, 
characterized by ${\bm M}$ along one of the $\langle 100\rangle$  
crystallographic directions.~\cite{ross_PRX,savary_PRL,chang,applegate}
 In that case, applying
a field along [100] selects one ordered state and the system merely exhibits a crossover to
a paramagnetic state at a temperature of order 1 K. The entropy then displays  an
activated temperature dependence at low temperatures. In contrast, for small fields along [110] two
degenerate ordered states remain and for small fields along [111], three degenerate states
remain. Thus, there should continue to be a low temperature phase transition 
in sufficiently weak magnetic fields along either $[110]$ or $[111]$, with only below the transition
the entropy going to zero.
By its nature, NLC does not allow for our
 theoretical calculations to converge at low enough temperatures and magnetic fields to fully confirm this picture.
It would thus be interesting to invesgitate its validity via $C(T,h)$ experimental measurements.

To summarize, the rather large anisotropic exchange interactions at play in Yb$_2$Ti$_2$O$_7$ lead to
the  development of a collective paramagnetic state at $T<T_c^{\rm mf} \sim 4$ K
with $\theta_{\rm CW} \lesssim T_c^{\rm mf}$, being thus somewhat ``hidden'' and difficult to quantitatively
describe using standard textbook (e.g. RPA and other mean-field like) methods.
In this material, and perhaps in other candidate quantum spin ice materials,
the fit of bulk thermodynamic data such as $C(T,h)$ and $M(T,h)$ using NLC may provide a useful alternative.

\section{Discussion \& Conclusion}
\label{discussion}

In this paper we have compared the results from numerical linked-cluster (NLC) calculations
of the temparature and magnetic field dependent magnetization of Yb$_2$Ti$_2$O$_7$ with those
obtained from experimental measurements on this material in a temperature range $T\in [1.8,20]$ K and
magnetic field range $h\in [0.2,5]$ T. The NLC calculations were performed on a Hamiltonian describing the
interactions between pseudospins $S=1/2$ and characterized by effective anisotropic exchange couplings
determined via inelastic neutron scattering (INS) measurements in the polarized paramagnetic state of
Yb$_2$Ti$_2$O$_7$.~\cite{ross_PRX}
The overall agreement between the NLC and the experimental results were found to be excellent (see Fig.~\ref{fig:3panels}).
Conversely, NLC magnetization results obtained using the exchange couplings determined from a random phase approximation
analysis of the diffuse paramagnetic neutron scattering~\cite{chang,thompson_PRL,already_critiqued} 
were found to be in significant disagreement with the experimental magnetization measurements.
The excellent argreement between NLC and experimental measurements of $M(T,h)$ shows that
(i)  the proposed nearest-neighbor exchange model~\cite{ross_PRX} is quantitatively accurate to describe Yb$_2$Ti$_2$O$_7$, 
that 
(ii) high-field inelastic scattering in the polarized paramagnetic state is a reliable method for extracting 
effective exchange parameters and that
(iii) long-range
dipolar interactions appear to not play an important role in the energetics of the system.
The latter conclusion is perhaps a bit surprising given that long-range dipolar interactions play
a key role in the physics of classical (dipolar) spin ices.~\cite{denHertog,gingras_CJP,isakov_PRL}
One obvious difference is that, relative to the $J_{zz}$ effective exchange interaction, the nearest-neighbor
contribution of the magnetostatic dipole-dipole interactions in Yb$_2$Ti$_2$O$_7$ is approximately one order of magnitude smaller than
in the classical dipolar spin ices.~\cite{bramwell_PRL2001,wiebe_prl,denHertog,yavorskii}
At the very least, one may consider that the nearest-neighbor contribution of the dipolar interactions are incorporated in 
the $\{ J_e \}$ effective anisotropic exchange. Then, it thus appears that the perturbative long-range (beyond nearest-neighbor) 
part of the dipolar interactions plays no dramatic role in a field greater than 0.2 T and down to 1.8 K (this work) or, in zero field,
 down to  approximately 0.7 K as found in Ref.~[\onlinecite{applegate}]. It may be that,  if Yb$_2$Ti$_2$O$_7$  does possess
a ferrimagnetically ordered state with a magnetization along of the  $\langle 100 \rangle$ axes, 
that the sole role of dipolar interactions is
to weakly renormalize the critical temperature and the level of quantum fluctuations, along with inducing ferrimagnetic domains. 
Yet, perhaps one should not be too expedient in assuming a generic irrelevance of 
long-range dipolar interactions of order $10^{-1}$ compared to $J_{zz}$
for any candidate quantum spin ice material. 
One can imagine that dipolar interactions may play an important role, possibly inducing novel phases, 
in a material that would, based solely on its effective
anisotropic exchange couplings $\{ J_e \}$, find itself at the boundary between 
the various semi-classical and intrinsically quantum phases identified in mean-field lattice
gauge theories of quantum spin ices.~\cite{savary_PRL,lee_onoda}


We believe that it would be, at this time, very interesting and most useful to carry out similar studies that
combine inelastic neutron scattering and thermodynamic bulk measurements for other candidate quantum spin ice materials, with
Yb$_2$Sn$_2$O$_7$ and Yb$_2$Zr$_2$O$_7$ being obvious choices.
From the lessons learned in the present work, as well as from Ref.~[\onlinecite{applegate}], 
we anticipate that NLC will contribute to developing a quantative parametrization of microscopic models of
quantum spin ices and help pave the way for an ultimate understanding of these fascinating systems. 
However, for this program to be successful,  the availability of
large high resolution inelastic neutron scattering data sets will likely prove to be essential.
In parallel, in-field single crystal measurements which properly account for 
demagnetization effects will also be necessary. 
To end on Yb$_2$Ti$_2$O$_7$, we believe that,
notwithstanding the evidence presented here and elsewhere~\cite{yasui,chang}, it is by no
means certain~\cite{hodges,but_maybe_not} at this time that long-range ferrimagnetic order develops below 0.26 K
 ~\cite{ross_YbTO_stuff}
in this very interesting material.
More experiments on well-characterized samples are needed to settle this question.



\begin{acknowledgements}
We acknowledge useful discussions with  P. dalmas de R\'eotier, 
B. Javanparast, P. McClarty, J. Thompson and A. Yaouanc.
This work is supported in part by NSF grant number DMR-1004231, the NSERC of Canada
and the Canada Research Chair program (M.J.P.G., Tier 1) and by the Perimeter Institute for Theoretical
Physics. Research at the Perimeter Institute is supported by the Government of Canada through Industry 
Canada and by the Province of Ontario through the Ministry of Economic Development \& Innovation.
\end{acknowledgements}


\begin{thebibliography}{99}
\bibitem{what_pm_and_z_means}
The pyrochlore lattice can be described by a FCC space lattice with a four-site (tetrahedron) sublattice basis.
The $\hat z$ quantization axis is along the local [111] cubic direction at each of the four
sublattice site, and $\pm$ (or equivalently, $x$ and $y$) refers to the two orthogonal local directions.
\bibitem{g-tensor}
The ``transverse'' (to the local [111] direction \cite{what_pm_and_z_means})
$\langle \psi^\pm J^\pm  H_{\rm int} \vert \psi^\mp \rangle$ matrix elements determine the Zeeman energy splitting
for an hypothetical magnetic field applied perpendicular to the local $[111]$ at a given site and which
 determines the $g_{xy}$ component of the $g$-tensor.
\bibitem{already_critiqued} The quantitative failure of a quantum spin ice Hamiltonian ${\cal H}_{\rm QSI}$ with
anisotropic exchange couplings obtained from a random phase approximation (RPA) fit of diffused neutron scattering,
 as reported in both Ref.~[\onlinecite{chang}] and Ref.~[\onlinecite{thompson_PRL}], to describe the zero field magnetic
specific heat had already been pointed out in Ref.~[\onlinecite{applegate}].
\end{thebibliography}

\end{document}